\documentclass[%
reprint,
superscriptaddress,
 amsmath,amssymb,
 aps,
 prb,
]{revtex4-1}

\usepackage{graphics}
\usepackage{graphicx}
\usepackage{yhmath}
\usepackage{color}
\usepackage{hyperref}
\definecolor{darkblue}{rgb}{0,0,.5}
\hypersetup{colorlinks=true, breaklinks=true, filecolor=blue, citecolor=blue, linkcolor=darkblue, menucolor=darkblue, urlcolor=blue}
\usepackage[normalem]{ulem}
\begin{document}

\title{Dual nature of localized phase defects in the In/Si(111) atomic wire array: \\impurities and short topological solitons}

\author{Abdus Samad Razzaq}
\affiliation{Interface Chemistry and Surface Engineering Department, Max-Planck-Institut f\"ur Eisenforschung GmbH, Max-Planck-Stra\ss e 1, 40237 D\"usseldorf, Germany} 

\author{Sun Kyu Song}
\affiliation{Center for Artificial Low Dimensional Electronic Systems, Institute for Basic Science (IBS), Pohang 37673, Korea}

\author{Tae-Hwan Kim}
\affiliation{Department of Physics, Pohang University of Science and Technology (POSTECH), Pohang 37673, Korea} 

\author{Han Woong Yeom}
\affiliation{Center for Artificial Low Dimensional Electronic Systems, Institute for Basic Science (IBS), Pohang 37673, Korea}
\affiliation{Department of Physics, Pohang University of Science and Technology (POSTECH), Pohang 37673, Korea} 

\author{Stefan Wippermann}
\affiliation{Interface Chemistry and Surface Engineering Department, Max-Planck-Institut f\"ur Eisenforschung GmbH, Max-Planck-Stra\ss e 1, 40237 D\"usseldorf, Germany}

\date{\today}

\begin{abstract}

We demonstrate the existence of atomically-sized topological solitons in a quasi one-dimensional charge density wave system: indium atomic wires on Si(111). Performing joint scanning tunneling microscopy and density-functional calculations, we show that the Si(111)-(8$\times$2)In surface features two conceptually different types of abrupt phase flip structures. One is caused by In adatoms and is, hence, non-solitonic in nature. The other one is an abrupt left-chiral soliton.

\end{abstract}

\maketitle

\section{Introduction}

Topological solitons -- local solitary wavepackets that connect two distinct but energetically degenerate states -- play an important role across many branches of science, including hydrodynamics, fiber optics, biophysics, quantum field theory, cosmology, materials science and condensed matter physics. In the context of condensed matter, screw dislocations in crystals are a commonly observed example for such a solitonic, topological defect. Topological defects, however, are known to occur not only within the atomistic structural properties of materials, but also within their electronic structure. Atomic-scale quasi one-dimensional (1D) electronic materials have received considerable attention in this context. They are of interest for potential applications in emerging nano-scale electronic devices \cite{chen,lu} and display exotic properties originating from enhanced many body interactions \cite{bockrath}. Utilizing topological solitons as carriers of charge or information in quasi-1D electronic systems is expected to lead to novel device functionality, e.g., robust single electron transistors and novel types of information processing.

In quasi 1D electronic materials, topological solitons were first described in polyacetylene \cite{heeger1,heeger2}. Here, they take the form of phase defects in the pattern of alternating single and double carbon bonds within the polymer chain. These solitonic phase defects carry charge and were shown to be highly mobile. Since their formation energy is less than the energy of the bandgap, they contribute significantly to the electron transport properties in these materials. In order to understand soliton formation and dynamics in electronic materials in detail, surfaces of low dimensional electronic materials and especially surface reconstructions with quasi 1D electronic properties are suitable model systems.

\begin{figure}[b]
  \centering
    \includegraphics[width=0.48\textwidth]{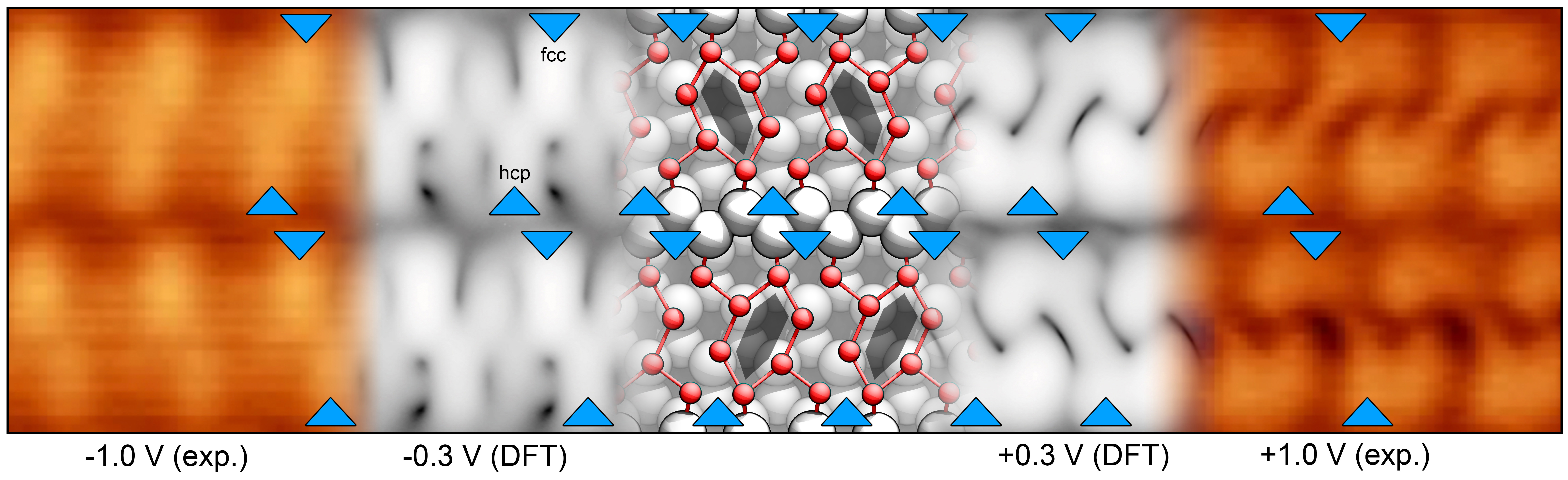}
\caption{\label{struct_stm} (Color online) Measured (at $\pm 1.0$ V) and simulated (at $\pm 0.3$ V) filled and empty state STM images for the pristine (8$\times$2) hexagon structural model. Blue triangles and grey hexagons mark the positions of the In trimers and In hexagons, respectively. Red and white balls represent indium and silicon atoms, respectively.}
\end{figure}

In this context, the ordered atomic wire array formed at the Si(111)-(4$\times$1)/(8$\times$2)In surface \cite{yeom99} is a particularly popular model system. It features a triple-band Peierls transition into a quasi one-dimensional (1D) charge density wave (CDW) ordered ground state with (8$\times$2) translational symmetry \cite{wipper10,kim20}. Solitons have been intensively investigated in this system \cite{kim12,zhang}. They appear as various long and short defects, where the CDW phase flips or slips along the wire direction across the position of the defect \cite{cheon}. The long extended phase flip (PFD) and phase slip (PSD) defects are generally considered to be intrinsic solitons of the system. Due to the presence of a band inversion in this system, cf. Fig. 5 in Ref. \cite{natrans}, these solitons exhibit chiral topological properties \cite{cheon}. Chiral switching processes of their topology were recently shown to comprise an Abelian group enabling robust computations with solitons as information carriers \cite{kim17}. The nature of the short PFDs and PSDs, on the other hand, has been discussed controversially. At present, they are generally considered to be of non-solitonic nature, but caused by In adatoms instead. Moreover, these adatom related defects are used intentionally to trap chiral topological solitons for STM imaging \cite{cheon} and are considered an essential circuit element in information processing schemes \cite{intertwined}.

In the present work, we perform joint STM measurements and density-functional theory (DFT) calculations to derive atomistic models of the most commonly observed phase defects in the Si(111)-(8$\times$2)In CDW phase. We demonstrate that the short PFD exists in \emph{both} solitonic and non-solitonic form. The adatom-induced and solitonic defects are structurally almost identical and closely resemble each other in scanning tunneling microscopy (STM) experiments. This is likely the origin of the long-standing debate about the origin of the short PFD/PSD. The DFT calculations also predict an attractive interaction between In adatom-induced and solitonic phase defects. This attractive interaction gives rise to the formation of a composite defect, which in its shortest possible form is the short PSD.

\section{Experiments and Calculations}

The STM experiments were carried out in an ultrahigh-vacuum ($8\times10^{-9}$~Pa) low-temperature STM (Unisoku Co., Ltd., Japan). The self-assembled In atomic wires on Si(111)-(4$\times$1)In surface were prepared by evaporating one monolayer of In onto the clean Si(111)-(7$\times$7) surface at an elevated temperature \cite{yeom99,kim17}. For STM measurements, the In atomic wires were cooled down well below the CDW transition temperature (either 4 or 78 K), which leads to the Si(111)-(8$\times$2)In CDW phase. In order to observe the same defect for a long time without thermal drift, we used a commercial temperature controller (Lake Shore Cryotronics, Inc., USA) to keep the sample temperature within $\pm$1 mK. All STM images presented here were taken in the constant-current mode with an electrochemically etched tungsten tip. We obtained filled and empty state STM images at the same place by changing the sample bias of $\pm$1 V alternatively while maintaining the tunnelling current of 50 pA.

We performed DFT calculations within the local density approximation (LDA) \cite{Ceperley_LDA} as implemented in the Vienna \emph{Ab initio} Simulation Package (VASP) \cite{Kresse-VASP}. We follow Stekolnikov {\em et al.} \cite{prl2007} concerning the numerical details. In $4d$ electrons were treated as core electrons. The surface was modeled using three Si bilayers. A (8$\times$21) surface unit cell was used, apart from the adsorption energy surface calculations which were performed using a (8$\times$12) lateral periodicity. The Brillouin zone (BZ) integrations in the electronic structure calculations were performed using uniform meshes equivalent to 256 points within a (1$\times$1) unit cell.

\section{Results and Discussion}

Fig. \ref{struct_stm} shows the hexagon structural model of the low temperature (8$\times$2) phase overlaid with simulated and measured STM images, respectively. The structure consists of an alternating arrangement of In wires and Si Seiwatz chains. Each In wire is comprised of two individual In zigzag chains. During CDW formation, regularly spaced trimers are formed within each individual In zigzag chain, as marked by blue triangles in Fig. \ref{struct_stm}. The trimerization introduces a doubling of the lateral periodicity along the wire direction from (4$\times$1) to (4$\times$2) translational symmetry. Note that within a single (4$\times$2) unit cell, the trimers can be placed in four different arrangements. Introducing alternating trimer placements between neighbouring (4$\times$2) wires, the periodicity perpendicular to the wire direction can be extended to (8$\times$2) or (16$\times$2) translational symmetries \cite{kim20}. Subsequently, a shear movement of the trimerized In zigzag chains against each other completes the transition into the CDW state \cite{wipper10,ortega}.

\begin{figure}[t]
  \centering
    \includegraphics[width=0.48\textwidth]{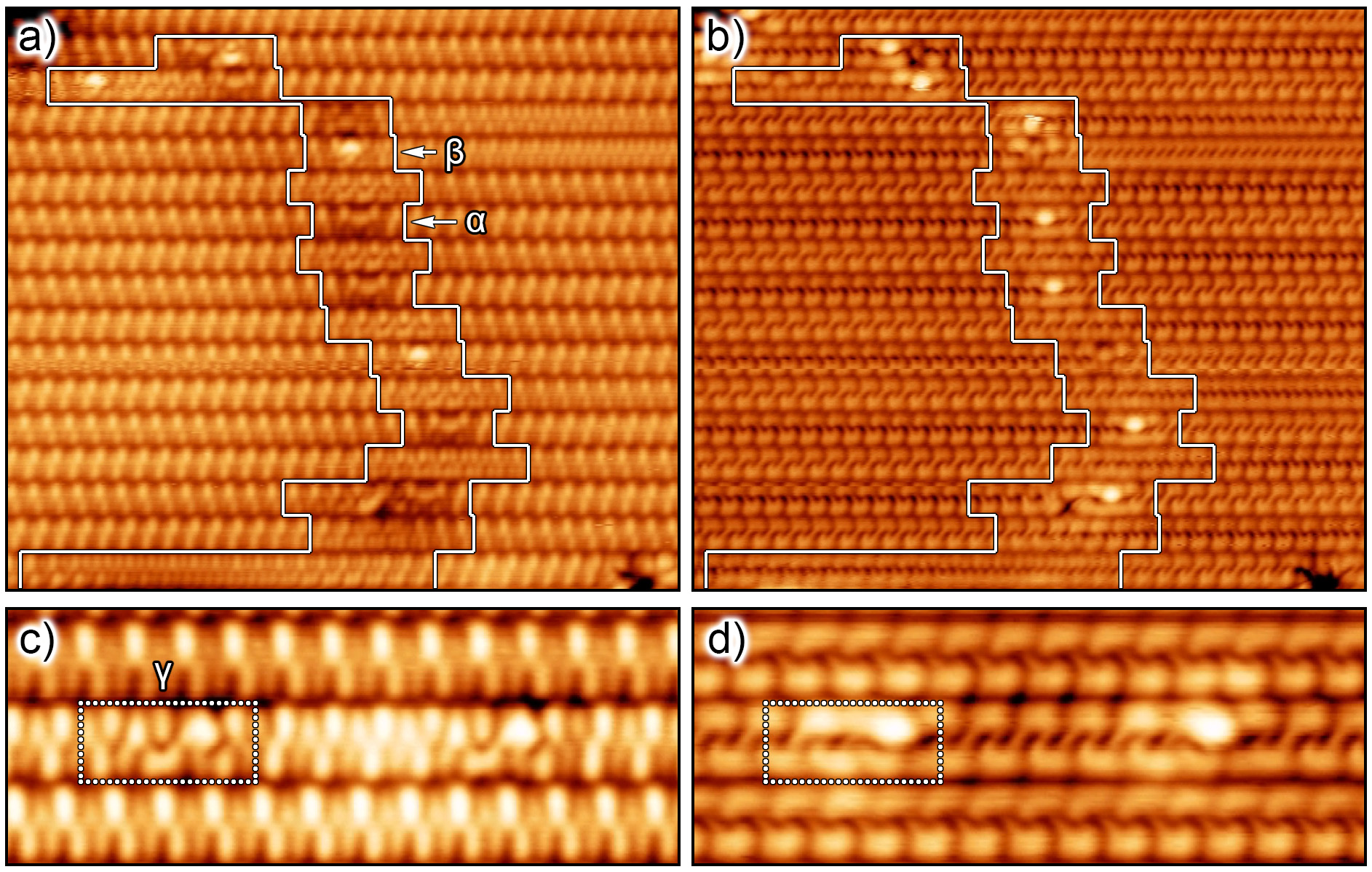}
\caption{\label{stm_exp} (Color online) \textbf{a/c}) Filled and \textbf{b/d}) empty state STM images at $T = 78$ K. All STM images were taken on the same area with a tunneling bias $V_S = \pm 1.0$ V, respectively, and a tunneling current $I_t = 50$ pA. }
\end{figure}

The filled state STM image shows regularly arranged rows of wires consisting of bright protrusions spaced at multiples of 2 of the $\times1$ lattice constant $a_0$. These protrusions correspond to the position of the aforementioned In trimers. The spaces between the trimers appear significantly darker in the filled state image. This trend reverses in the empty state image, where the trimers are darker and the space between trimers is now bright. Both simulated images are in excellent agreement with the STM measurements. Note that the simulated images are computed at a somewhat lower bias of $\pm 0.3$ V compared to $\pm 1.0$ V used in the experiments, to account for the DFT band gap underestimation. 

In the simulated STM images, the trimers on the upper of the two In zigzag chains appear slightly brighter than the trimers on the lower chain. This behaviour is attributed to the upper chain trimers being located on fcc hollow sites, as opposed to the lower ones located on hcp hollow sites of the substrate. Such a difference in intensity is also observed consistently in the STM measurements, cf. Fig. \ref{struct_stm} for a direct comparison. The intensity difference allows us to determine directly from the STM measurements the spatial orientation of the substrate and the precise positioning of the In trimers. This information is crucial when modelling the structure and chirality of solitons. The In chains alone are rotationally invariant and, hence, so are some of the possible solitons. The substrate breaks rotational symmetry and thereby provides a natural point of reference to distinguish between different types of solitons.

It is also apparent in Fig. \ref{struct_stm} that the individual (4$\times$2) hexagons constituting the (8$\times$2) reconstruction can be oriented in two different directions. Moreover, the (4$\times$2) hexagons can be shifted by a single $\times$1 lattice constant $a_0$ along the wire direction. Note that these orientation and phase states of the CDW correspond to specific placements of the trimers within the respective (4$\times$2) unit cell. In consequence, these degrees of freedom in the CDW orientation and phase give rise to a 4-fold degeneracy of the CDW ground state \cite{cheon,sol}. Forming a phase boundary between regions with different CDW phases and orientations then leads to the formation of phase slip defects (PSDs) and phase flip defects (PFDs), respectively.

Figs. \ref{stm_exp}a and \ref{stm_exp}b show filled and empty state STM images, respectively, of characteristic PFDs and PSDs on the (8$\times$2) surface. The STM measurements reveal two visually different types of abrupt phase flip defects, labeled $\alpha$ and $\beta$ in Fig. \ref{stm_exp}a. Structurally, they are very similar in appearance. However, the $\alpha$ defect is bright in the empty state image, whereas the $\beta$ defect appears bright in the filled state image. Both defects feature an intact row of trimers in the lower In chain, but feature a phase slip in the upper In chain, corresponding to a distance of $3 a_0$ between two adjacent trimers at the defect site. According to the intensity difference of the trimers and comparing to Fig. \ref{struct_stm}, the phase slip occurs in the trimer row located on the fcc lattice sites. The combination of one pristine row of trimers and one trimer row containing a phase slip defect then leads to the overall formation of a phase flip of the whole In hexagon chain.

In the terminology of Ref. \cite{cheon}, both the $\alpha$ and the $\beta$ defect exhibit a left-chiral topology. The defects are part of an extended domain wall perpendicular to the wire direction, as marked by the enclosed region in Fig. \ref{stm_exp}. In the neighbouring wire located between the short $\alpha$- and $\beta$-PFDs we observe an extended PFD, which is a left-chiral soliton. This pattern of alternating extended left-chiral solitons and left-chiral short $\alpha$- or $\beta$-PFDs continues with surprisingly long range order across multiple In wires.

\begin{figure}[t]
  \centering
    \includegraphics[width=0.35\textwidth]{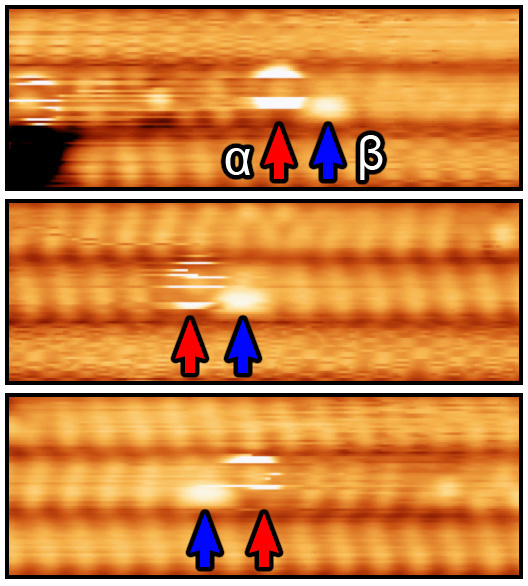}
\caption{\label{switch} (Color online) Filled state STM images of different surface areas at $T = 4$ K, $V_S = -1.5$ V. The increased tunneling bias induces switching of $\alpha$-PFDs. In contrast, $\beta$-PFDs remain static.}
\end{figure}

A third type of defect commonly observed in our measurements is shown enlarged in Figs. \ref{stm_exp}c and \ref{stm_exp}d, labeled $\gamma$. This defect apparently consists of a $\alpha$-PFD and an additional phase slip in the lower trimer row. In total, a phase slip in the whole In hexagon chain is obtained, resulting in a short PSD. This short PSD is again in marked difference compared to its extended counterpart. The extended phase defects are unambiguously considered to be genuine solitons \cite{cheon}, whereas the the nature of the short ones remains controversial \cite{zhang}.

\begin{figure*}[t]
  \centering
    \includegraphics[width=0.9\textwidth]{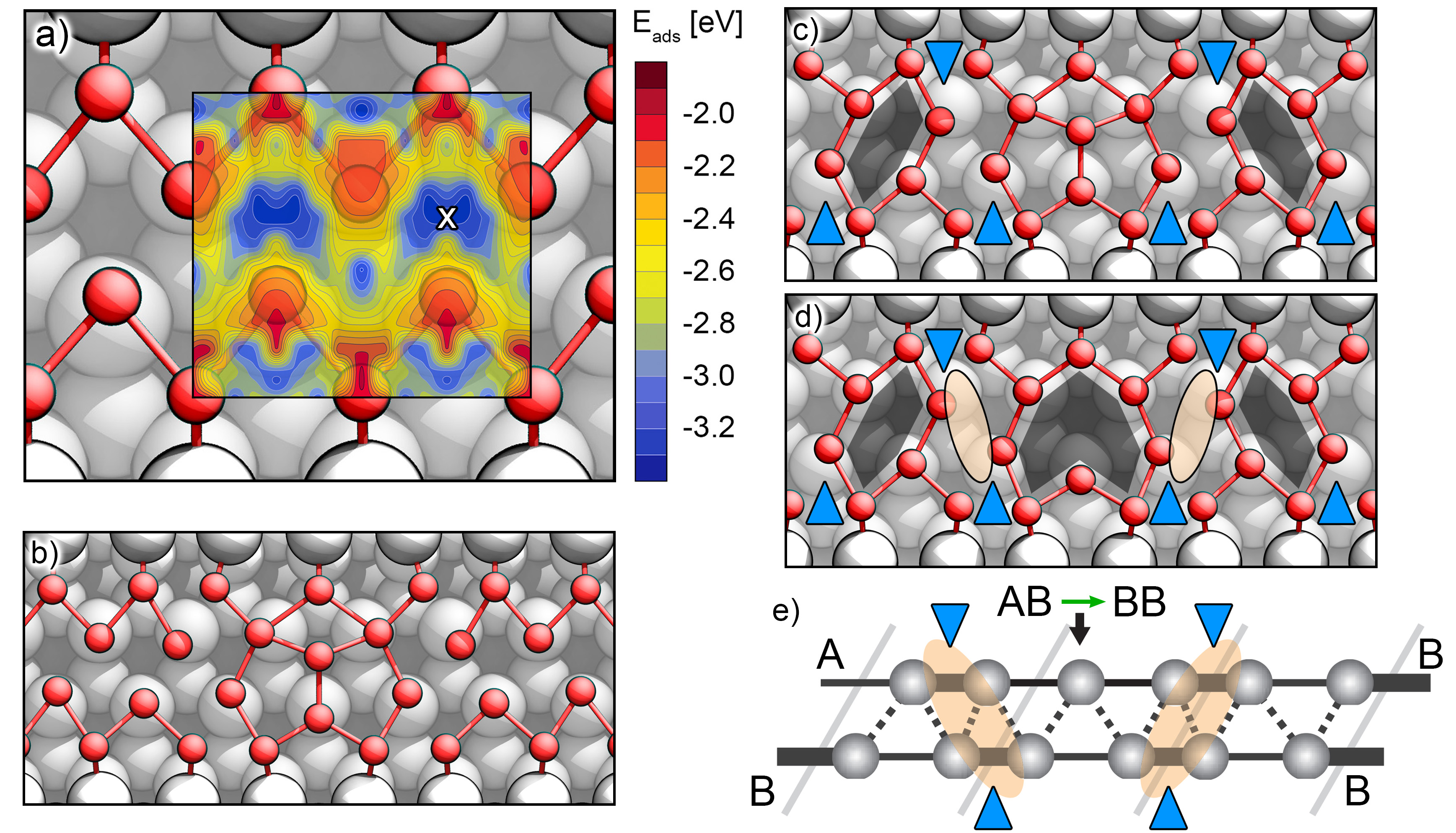}
\caption{\label{aes} (Color online) \textbf{a)} Adsorption energy surface (AES) calculated for an indium adatom on the Si(111)-(4$\times$1)In surface overlaid on the (4$\times$1)In structure model. The color scale shows the adsorption energy. The minimum energy adsorption site is marked by a white cross. \textbf{b/c)} Top view of the relaxed atomistic model with the In adatom adsorbed at the minimum energy intrachain site on the (4$\times$1) and (8$\times$2) surfaces, respectively. \textbf{d)} Top view of the relaxed solitonic phase flip boundary. Red and white balls in the structure models represent indium and silicon atoms, respectively. Blue triangles mark the positions of In trimers, see text. Ovals indicate the positions of the bright protrusions observed in STM. \textbf{e)} Schematic representation of a left chiral soliton within the coupled double Peierls chain model, derived from SSH theory. Reproduced with permission from Ref. \cite{cheon}.}
\end{figure*}

Previous studies strongly suggested that these short PFDs and PSDs are caused by In adatoms \cite{kim12,intertwined}. In line with their findings, we observe that with increasing In dosage the number of $\alpha$-PFDs rises in proportion. The number of $\beta$-PFDs increases as well. However, their number appears not to be directly correlated with the In dosage, indicating they are created by a secondary process. In order to demonstrate that the $\alpha$- and $\beta$-PFDs are conceptually different, despite their structural similarity, we performed STM measurements at a relatively high sample bias of $V_S = -1.5$ V  because Zhang \textit{et al}. revealed that In adatoms could hop along the wires due to high sample biases \cite{zhang}. Interestingly, we observed a rapid switching of the $\alpha$-PFDs, where the defect alternates between either its regular appearance or a very bright feature in its place, cf. Fig. \ref{switch}. Note that above 100 K, the very bright feature becomes highly mobile along the wire direction. The $\beta$-PFD, in contrast, remains completely inert even with the high sample bias. We propose that the $\alpha$-PFD is indeed created by In adatoms. The switching event is the ejection/injection of the In adatom into the PFD, where the very bright signal on top of the $\alpha$-PFD is the ejected In adatom. Since the $\beta$-PFD does not display any switching behaviour, we conclude that it does not contain any In adatoms. Its overall structure is apparently identical, just without the additional In adatom.

As a first step towards establishing atomistic structural models of these defects, we searched for lateral adsorption sites of an In adatom by calculating the adsorption energy surface (AES). The lateral position-dependent adsorption energies for a single In adatom per (8$\times$12) surface unit cell are shown in Fig. \ref{aes}a. The results differ slightly from previous AES calculations \cite{wipper08} performed in a smaller (4$\times$3) unit cell, leading to adatom-adatom interactions caused by the periodic boundary conditions. In the AES, both fcc and hcp interchain positions correspond to energy minima (maximizing the energy gain by adsorption), with the fcc site being the lowest energy configuration identified and marked by a white X in Fig. \ref{aes}a. This is consistent with the observation that in the STM experiments the In adatom is preferentially found on the upper In chain on a fcc lattice site. Intrachain adsorption represents another local energy minimum but leads to significantly higher energies and is thus energetically less favourable.

Consecutively, we placed an In adatom on these adsorption sites and relaxed the surface structure without any constraint. We note that for the same lateral adsorption site two different energy minima were found: (i) the In adatom is adsorbed on top of the ideal (4$\times$1) In wire structure, barely distorting the (4$\times$1) reconstruction at all, and (ii) the In adatom is located in-plane with the (4$\times$1) wires. The latter configuration is shown in Fig. \ref{aes}b. In adatoms on the (4$\times$1) surface are generally assumed to be highly mobile \cite{intertwined}. This is in line with our observation that the very bright feature in Fig. \ref{switch} becomes mobile above 100 K. We expect that these mobile In adatoms correspond to configuration (i), whereas configuration (ii) is 0.54 eV deeper in energy compared to (i) and is the global energy minimum on the (4$\times$1) surface. Note that the AES in Fig. \ref{aes} always reports the lowest energy minima, as shown e.g. in Fig. \ref{aes}b.

Interestingly, the adsorption structure shown in Fig. \ref{aes}b induces local distortions of the ideal (4$\times$1) wire that coincide with the trimerization and shear modes known \cite{wipper10} to transform the (4$\times$1) into the (8$\times$2) reconstruction. For that reason, the same In adatom induced local wire structure represents an energy minimum on both the (4$\times$1) room temperature and (8$\times$2) low temperature structures, cf. also Ref.  \cite{dynamical} in this context. Figure \ref{aes}c shows the globally most stable surface structure that we identified. Consistent with the STM images in Fig \ref{stm_exp}a, the trimer chain in the lower wire is uninterrupted, whereas a phase slip of a single lattice constant $a_0$ is introduced in the upper trimer chain.

\begin{table}[b]
\centering
\begin{tabular}{l|cc}
\hline\hline
 & \multicolumn{2}{c}{\textsc{Energy of Adsorption/Formation [eV]}} \\
 & (4$\times$1) ideal & (4$\times$2) hexagon \\
\hline
($\alpha$) In adatom & $-$3.57/$-$3.48 (hcp/fcc) & $-$3.34 \\
($\beta$) heart & 0.07/0.17 (hcp/fcc) & 0.31 [0] / 0.08 [+2] \\
\hline\hline
\end{tabular}
\caption{\label{tabE} Calculated energies of adsorption/formation for different defect types. Note that defect formation on hcp/fcc lattice sites can only be distinguished on the (4$\times$1) surface. On the (4$\times$2) surface periodic boundary conditions enforce the presence of two defects, one on the hcp and one on the fcc site. Thus the reported (4$\times$2) energies are averages. All calculations for the (4$\times$2) hexagon surface were performed in the (8$\times$21) unit cell, containing one (4$\times$2) wire separated by a frozen (4$\times$1) wire to minimize interchain coupling. Values in brackets indicate the defect charge state, see text.}
\end{table}

Tantalizingly, the local In adatom-induced structure in absence of the adatom itself closely resembles two corner-sharing In hexagons. In Fig. \ref{aes}c, the shared corner is the one located at the regular In atom located directly above the central In adatom. We removed the In adatom from the structure shown in Fig. \ref{aes}c and relaxed the structure again without constraints. The largest relaxations are observed at the In atoms directly adjacent to the now absent In adatom, and amount to less than 0.2 \AA\ on average. Figure \ref{aes}d shows the fully relaxed structure. The corner-sharing hexagon structure, hence, represents another local energy minimum and is termed ``heart''-shaped structure in the present work due to its characteristic shape. In the language of Ref. \cite{cheon}, this boundary between two degenerate CDW ground states constitutes a left-chiral soliton. In Fig. \ref{aes}e we compare the corresponding left-chiral soliton derived from SSH theory for the coupled double Peierls chain \cite{cheon} to the heart-shaped $\beta$-PFD.

We note that within SSH theory solitons always exhibit a characteristic length scale, cf. Refs. \cite{heeger1,heeger2}. The extended solitons present in the In/Si(111) system are generally assumed to correspond to the global minimum of the soliton energy as a function of their length. For this reason, the possible existence of atomically-sized solitons remains controversial. We point to the fact, however, that the original SSH work as well as e.g. Ref. \cite{cheon} assume a harmonic potential for the atomic displacements. In contrast, the real system displays strong anharmonicities. Due to bond-making and -breaking during the phase transition, the harmonic approximation breaks down already for a single (4$\times$2) hexagon wire. Interwire interaction introduces a further site-dependent anharmonic contribution, as strongly indicated by the domain-wall formation visible in Fig. \ref{stm_exp}. However, beyond the harmonic approximation, there is no longer a guarantee that the soliton energy as a function of length has only a single minimum.

We propose that the $\beta$-PFD corresponds to a local minimum within the soliton formation energy and expect it to be described correctly by the SSH concept, though with a generalized anharmonic potential for the atomic displacements. Topologically, it is identical to the AB $\rightarrow$ BB left-chiral soliton introduced in Ref. \cite{cheon}, cf. Fig. \ref{aes}e. Moreover, we remark that the topology of an edge state is independent of its lateral extension. For this reason, the $\beta$-PFD must display topological properties analogous to its extended counterpart.

In the neutral charge state, the formation energy of the $\beta$-PFD is obtained from DFT as 309 meV. We immediately point out, however, that the $\beta$-PFD is expected to assume a +2 positively charged state: SSH theory predicted \cite{cheon} that upon formation of a left chiral soliton two occupied electronic states are lifted above the Fermi level. Hence, two electrons are released to the reservoir. Depending on the Fermi level, we thus expect that the $\beta^{+2}$-PFD formation energy is less than 309 meV. This will be discussed further below. The formation energies are summarized in Tab. \ref{tabE}.

\begin{figure}[t]
  \centering
    \includegraphics[width=0.48\textwidth]{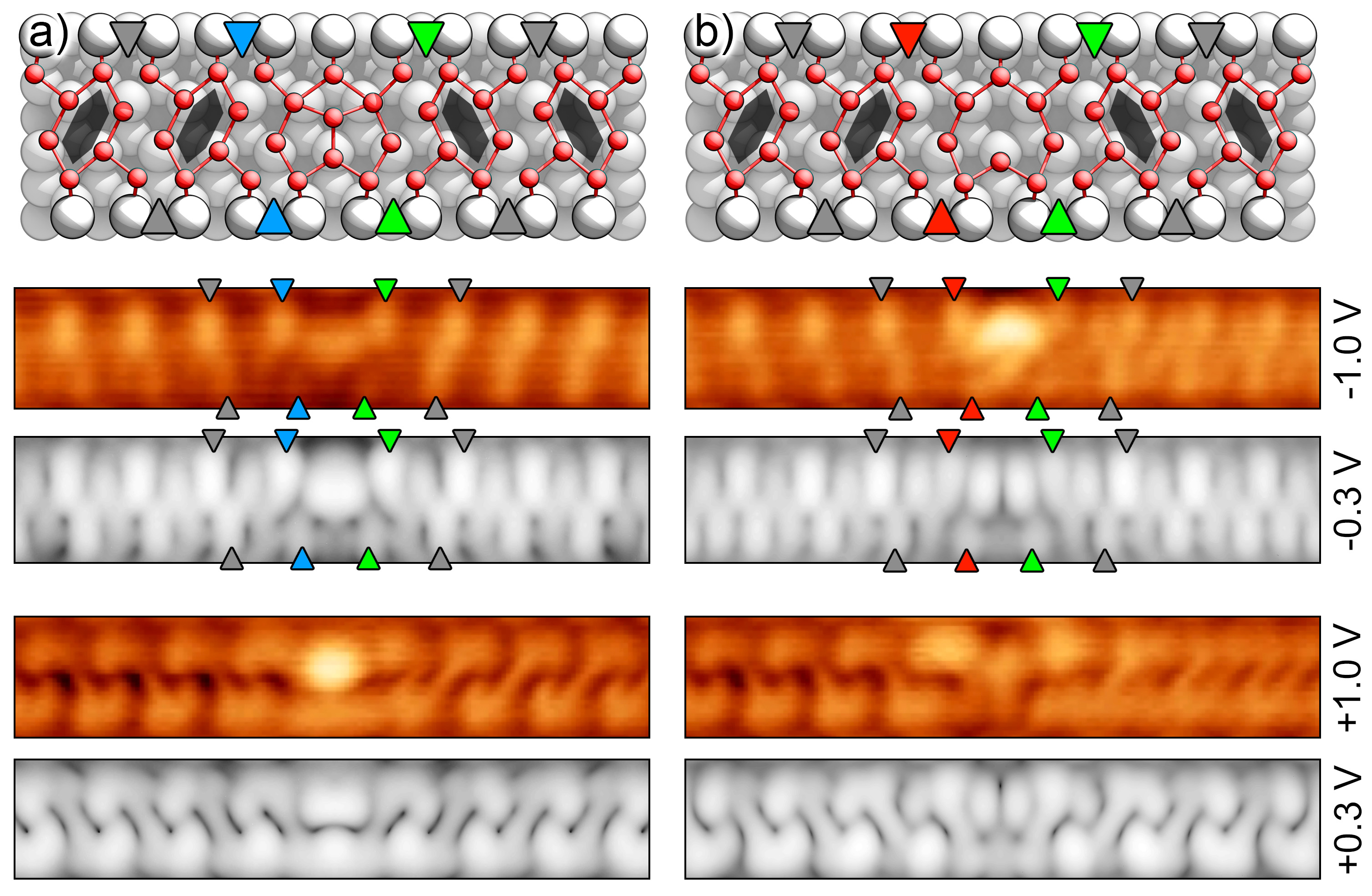}
\caption{\label{stmcomp} (Color online) High-resolution filled and empty state STM images in comparison to simulated STM images (in gray scale) at the indicated bias ($I_t = 50$ pA, $T = 78$ K). The simulated images are calculated at smaller bias potentials to compensate for the DFT band gap underestimation. Voltages in the simulations are specified with respect to the center of the CDW gap. The top row shows the relaxed structural models of the \textbf{a)} In adatom-induced defect and \textbf{b)} a solitonic phase boundary. Colored triangles and black hexagons mark the positions of the In trimers and In hexagons, respectively.} 
\end{figure}

\begin{figure*}[t]
  \centering
    \includegraphics[width=0.98\textwidth]{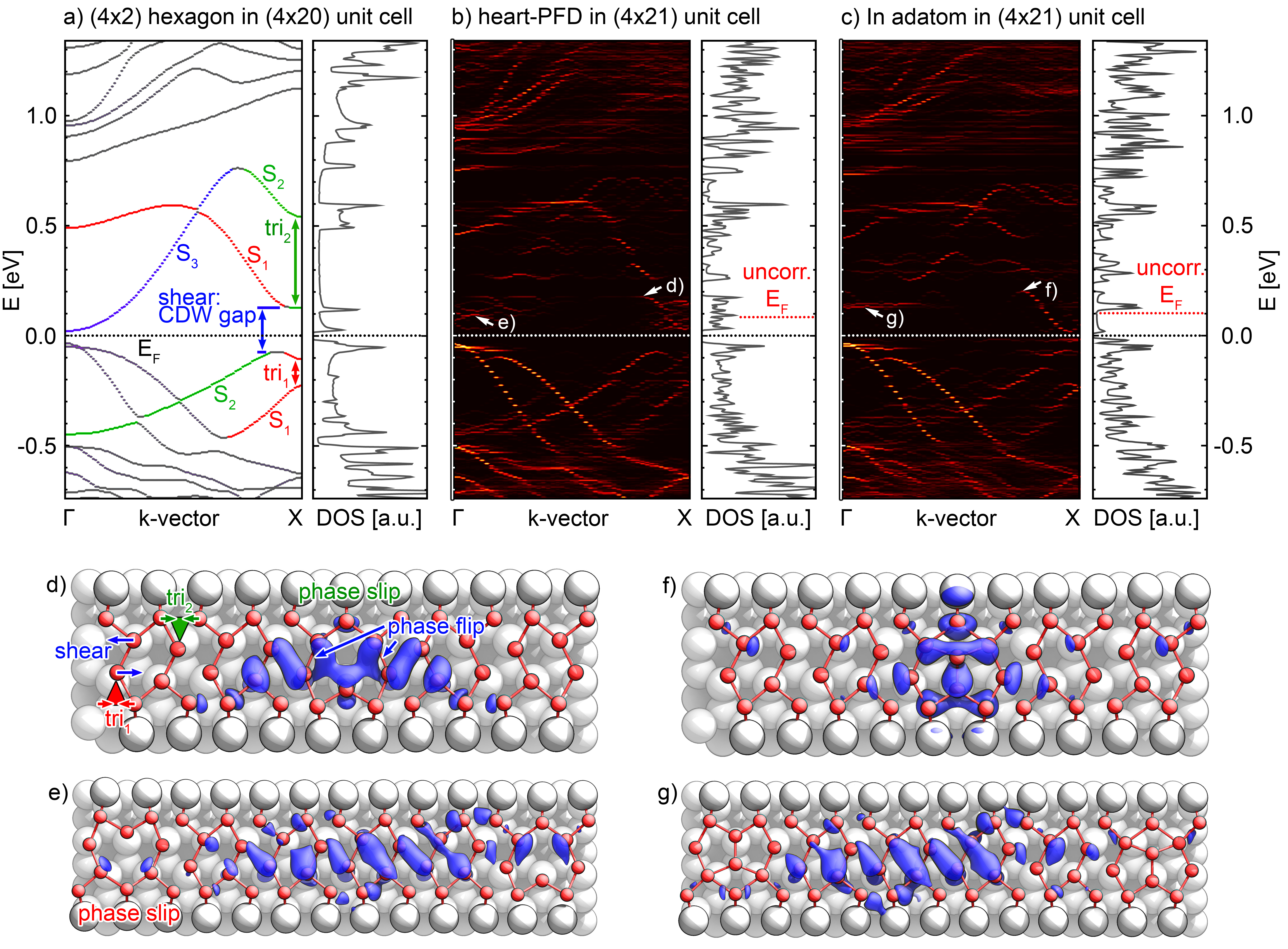}
\caption{\label{unfold} (Color online) \textbf{a-c)} Surface band structures and electronic densities of states along the $\Gamma$X-direction of the surface Brillouin zone. Bands of equal color are characterized by similar wavefunctions, cf. Fig. 5 in Ref. \cite{natrans} and Ref. \cite{cheon}. The band structures were unfolded within the (4$\times$2) surface Brillouin zone, see text. \textbf{d-g)} Isosurfaces (0.0015 e$^-$\AA$^{-3}$) of wavefunction square moduli for representative states, see text.}
\end{figure*}

We now turn to validating the predicted defect structures by comparing simulated STM images to the measurements reported in Fig. \ref{stm_exp}, as shown in Fig. \ref{stmcomp}. Colored triangles indicate the position of corresponding trimers in the structure models, as well as in the simulated and measured STM images. Comparing the high resolution STM data in Figs. \ref{stmcomp}a and b shows that indeed the overall size and local structure of both defects is identical. The adatom-induced defect, however, is bright in the empty state images and significantly darker in the filled state images, respectively. The heart-shaped PFD, on the other hand, shows exactly the opposite behaviour: it is bright in the filled state images and dark in the empty state images, respectively. Moreover, the In trimers are usually dark in the empty state images. For the heart-shaped PFD, in contrast, the trimers within the upper In zigzag chain that are located adjacent to the phase slip appear as bright features. This behaviour is well reproduced by our STM simulations.

We now turn to a discussion of the electronic properties. For the pristine hexagon reconstruction, these are commonly discussed in terms of the surface band structure. In the presence of defects, however, this is not directly possible due to the increased lateral size of the unit cell and the induced back-folding of the surface bands. In order to obtain an approximate band structure within the primitive surface Brillouin zone, we use the unfolding method introduced in Refs. \cite{bandup1,bandup2} by Medeiros \emph{et al.} Figure \ref{unfold}a shows the surface band structure of the (4$\times$2) hexagon obtained from unfolding the electronic states of the (4$\times$20) unit cell. As expected, the unfolded band structure is identical to the one obtained directly from the (4$\times$2) unit cell (not shown). In the next step, we then unfolded the band structures of the heart-shaped PFD and In adatom PFD, cf. Figs. \ref{unfold}b and \ref{unfold}c, respectively.

\begin{figure*}[t]
  \centering
    \includegraphics[width=0.98\textwidth]{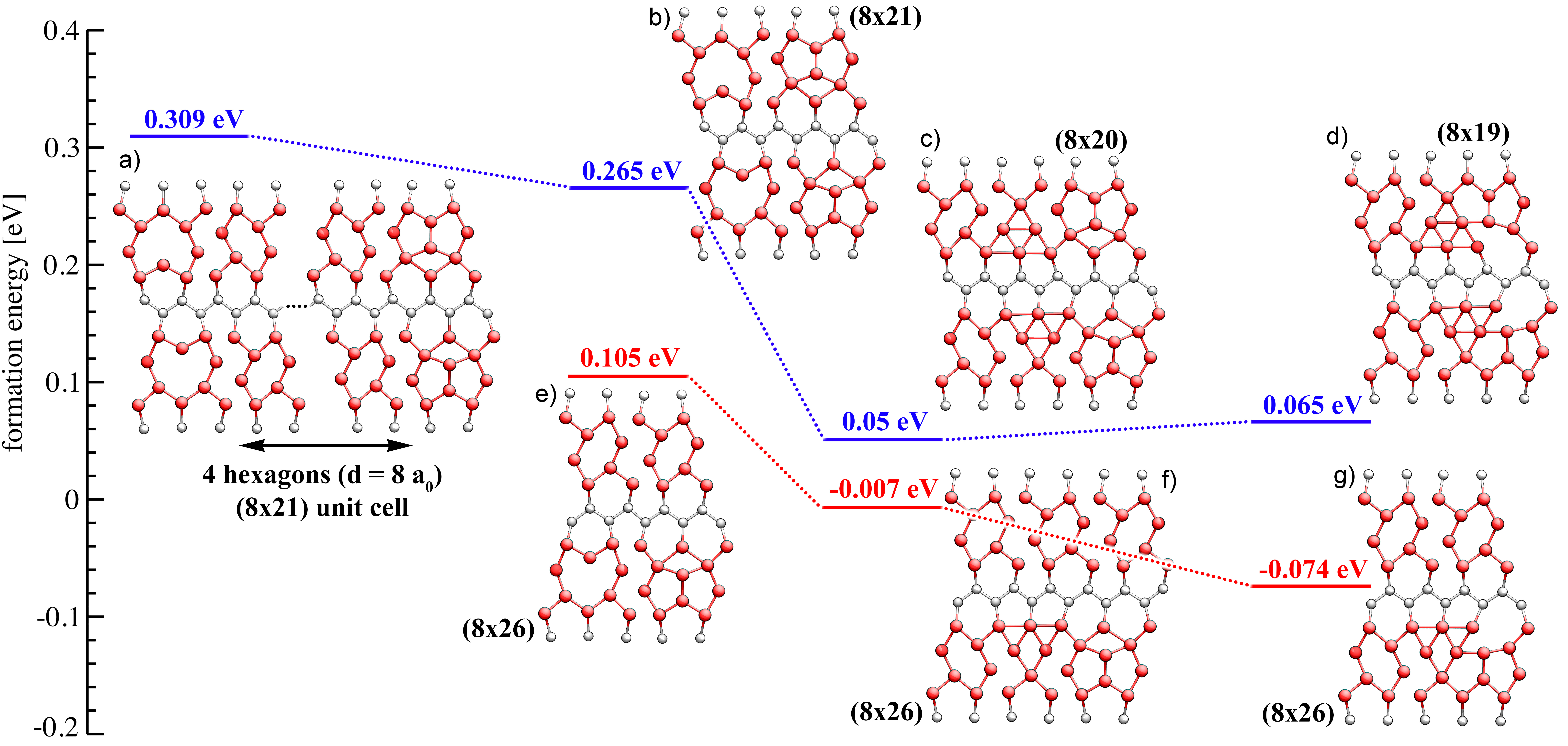}
\caption{\label{domain} (Color online) Formation energy of a single heart-shaped PFD as a function of distance to an In adatom PFD with and without interwire interaction. All formation energies are specified with respect to the pristine (4$\times$2) structure in place of the heart-shaped PFD.}
\end{figure*}

In order to discuss the impact of the PFDs on the electronic structure, it is instructive to revisit the mechanism of the (4$\times$2) hexagon formation. The (4$\times$1) wire consists of two metallic In zigzag chains. At low temperatures, each can undergo individually a period-doubling Peierls transition. Displacements of the outer In atoms, indicated by red and green arrows in Fig. \ref{unfold}d, lead to the formation of the aforementioned In trimers. In agreement with Peierls theory, gaps are opened in the corresponding S$_1$ and S$_2$ surface bands at the X-point of the (4$\times$2) surface Brillouin zone. One might expect the surface bands of both In zigzag chains to be degenerate. The coupling between the individual In zigzag chains, however, lifts the degeneracy between the S$_1$ and S$_2$ bands in analogy to the formation of bonding and antibonding states. As a result, the gap at the X-point in the S$_1$ band is located below the Fermi level, whereas the corresponding gap in the S$_2$ band is located above. Hence, the S$_1$ and S$_2$ bands cross each other before the X-point at 0.85$\overline{\Gamma\mathrm{X}}$. The actual CDW gap is then opened exactly at this crossing, cf. Fig. \ref{unfold}a: an additional shear displacement between the two zigzag chains, cf. blue arrows in Fig. \ref{unfold}d, leads to the formation of new interchain bonds between the two zigzag chains and thereby to the opening of the CDW gap. 

We note that the rehybridization that is responsible for opening the CDW gap at the S$_1$/S$_2$ crossing at the same time leads to a band inversion. It is precisely for this reason that this system is expected to feature interesting topological properties \cite{cheon,kim17}. Further details about the phase transition are discussed in Ref. \cite{wipper10}. An evolution of the band structure along the transition path and the formation of the band inversion are shown in Fig. 5 in Ref. \cite{natrans}.

We now turn to a discussion of the unfolded band structures in Fig. \ref{unfold}b/c. Overall, the unfolded bands closely resemble the (4$\times$2) band structure shown in Fig. \ref{unfold}a and the S$_1$ -- S$_3$ surface bands are still recognizeable. The S$_3$ band, however, that couples to the shear mode, is significantly broadened. This broadening is caused by the fact that the magnitude of the shearing is strongly modified in the vicinity of the PFD. At the center of the PFD, the direction of the shear motion is is inverted. The local variation of the shearing at different positions in the wire in turn leads to variations in the energetic position of the S$_3$ band.

An analogous broadening is observed for the S$_1$ and S$_2$ bands. Note in Fig. \ref{unfold}d that the trimer chain on the fcc site ($tri_2$) features a phase slip defect at the center of the PFD, leading to a variation in the energetic position of the S$_2$ band. The $tri_1$ chain on the other hand, is undisturbed in the same location. However, it is necessary to ensure periodic boundary conditions despite the flipping of the hexagon orientation. Therefore, the unit cell must contain a second PFD with opposite vertical orientation, cf. Fig. \ref{unfold}e. This mirrored PFD leaves the trimerization in the $tri_2$ chain unchanged, but introduces a corresponding phase slip defect in the $tri_1$ chain and hence broadening in S$_1$.

\begin{figure}[b]
  \centering
    \includegraphics[width=0.35\textwidth]{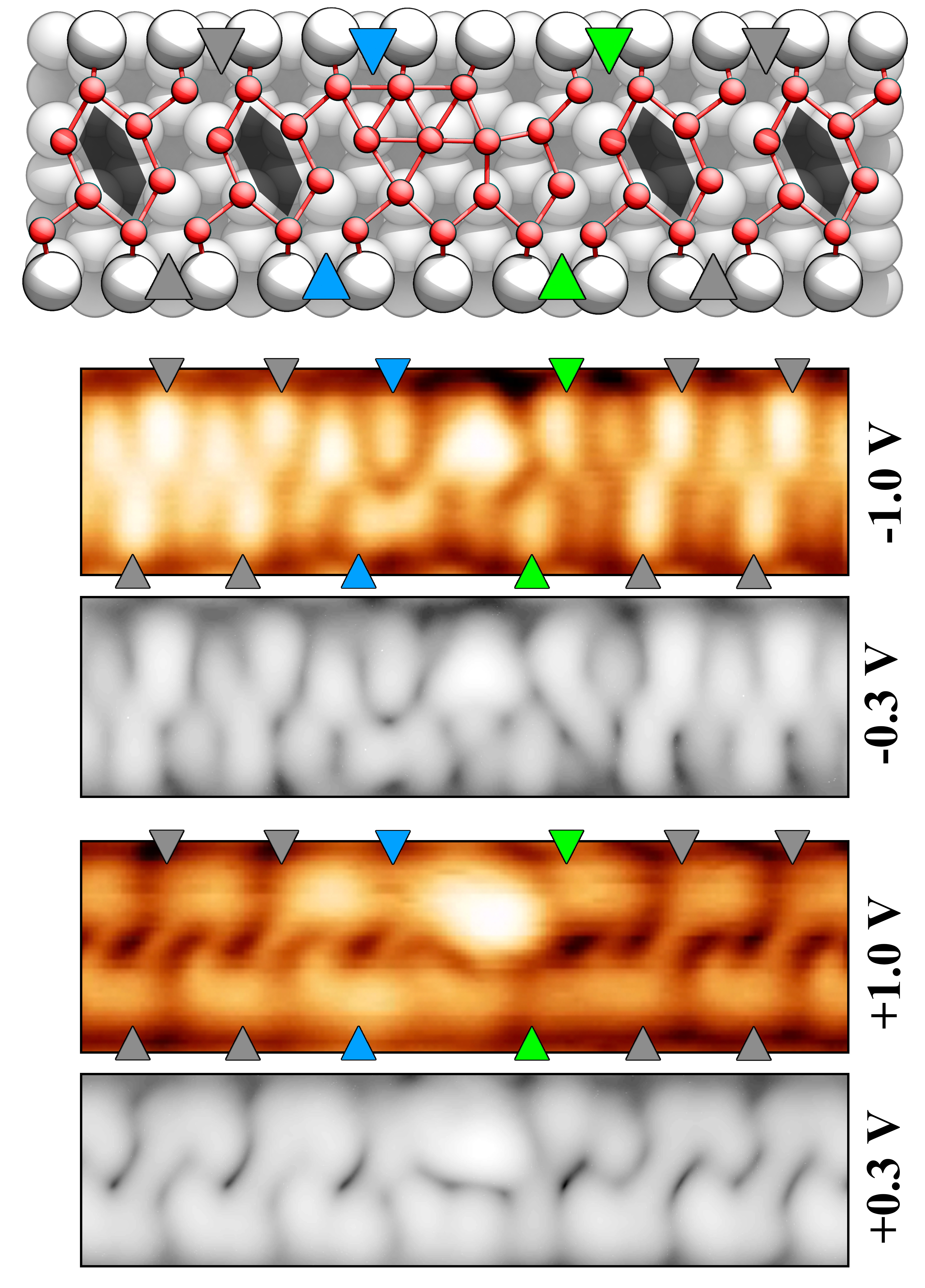}
\caption{\label{stmcomposite} (Color online) High-resolution filled and empty state STM images in comparison to simulated STM images (in gray scale) at the indicated bias ($I_t = 50$ pA, $T = 78$ K) for the $\gamma$-composite defect.}
\end{figure}

Furthermore, the surface bands split into a number of minibands separated by narrow gaps. This is a finite size effect: periodic boundary conditions are required in order to describe the surface, but introduce at the same time periodic images of the PFDs. Hence, our calculations describe finite wire segments enclosed between periodic PFDs. The periodic spacing of the PFDs modulates the wavefunctions in analogy to the well-known particle-in-a-box model. Figures \ref{unfold}e/g show isosurface plots for the wavefunction square moduli of representative states belonging to the S$_3$ band. With increasing unit cell size the gaps between the minibands decrease. The gap formation, however, is also of interest itself, e.g. in the context of multiple PFDs present within the same wire. Moreover, we note that the PFDs introduce a number of highly localized electronic states. Figures \ref{unfold}d/f show isosurface plots for the wavefunction square moduli of the lowest energy states localized to the respective PFDs.

Refs. \cite{cheon,kim17}, and more recently Ref. \cite{huda}, predicted from Su-Schrieffer-Heeger (SSH) theory \cite{heeger1,heeger2} that a left chiral soliton induces two additional electronic states above the Fermi energy. The formation of these new electronic states is accompanied by an exchange of charge with the reservoir at the Fermi level. This prediction was confirmed, also in Refs. \cite{cheon,kim17}, by scanning tunneling spectroscopy (STS) measurements. In fact, Erg\"un \emph{et al.} \cite{grand1} recently suggested that a correct description of solitons and the phase transition requires a treatment that is grand-canonical with respect to the charge. Our present DFT calculations, in contrast, are performed at a constant number of electrons. In this context, no exchange of charge with the reservoir is possible. Therefore, if the PFD shown in Figs. \ref{stmcomp}b and \ref{unfold}d is truly a left chiral soliton, instead of releasing its surplus charge into the substrate the charge must be accommodated in the lowest energy conduction states. Compared to the pristine (4$\times$2) surface, there must be one extra fully occupied band per heart-shaped PFD and a corresponding increase of the Fermi energy. Indeed, that is what we find.

Next to the band structures in Figs. \ref{unfold}a/b, we compare the electronic density of states (DOS) of the (4$\times$2) hexagon structure to the DOS of the heart-shaped PFD. The CDW gap remains open despite the presence of the PFD. The DFT Fermi level, however, is now shifted above the conduction band minimum. We verified from the calculated electronic occupations that the band filling of the conduction band is precisely 2 e$^-$ per PFD, in exact agreement with the prediction\cite{cheon} from SSH theory.

This observation now allows us to estimate the formation energy of the heart-shaped PFD in its +2 charge state. As indicated by the white/black dotted line in Fig. \ref{unfold}b, we set the Fermi energy to the same value as for the pristine hexagon structure. We then subtract the contribution to the band structure energy originating from the occupied states between this assumed value for $E_F$ and the uncorrected $E_F$ indicated by the red dotted line. This reduces the formation energy for the heart-shaped PFD from 309 meV in the charge neutral case to 80 meV in its +2 charge state.

Finally, we proceed to determine the structure of the $\gamma$ phase slip defect (PSD) shown in Fig. \ref{stm_exp}c/d. We hypothesize that the PSD is in fact a trapped left chiral soliton in the form of a heart-shaped PFD next to an In adatom PFD. As apparent in the first enclosed row in Fig. \ref{stm_exp}a/b, both defects can be placed next to each other, possibly suggesting an attractive interaction. Figure \ref{domain} shows the formation energy of a single heart-shaped PFD as a function of distance to a neighbouring In adatom PFD. Indeed, we find a slight attractive interaction and the formation energy is lowered by 40 meV for neighbouring PFDs compared to isolated PFDs (Fig. \ref{domain} a$\rightarrow$b). A much more significant energy lowering of 200 meV is achieved, however, by moving the heart-shaped PFD even closer by a single lattice constant $a_0$ (Fig. \ref{domain} b$\rightarrow$c). As a result, the heart-shaped PFD and the In adatom PFD no longer feature separate corners, but share a single corner in the center between the two defects. The structure of the In adatom PFD is sterically constrained by the presence of the adatom. It remains stable during structural relaxation. The heart-shaped PFD on the other hand evolves into a characteristic triangular structure (Fig. \ref{domain}c).

If the neighbouring In wire is in the pristine (4$\times$2) configuration, the energy can be lowered even further by moving the triangular structure another lattice constant $a_0$ towards the adatom PFD. Hence, the triangle and the In adatom PFD are now sharing a corner (Fig. \ref{domain} f$\rightarrow$g). We note that the structure obtained in this way is equivalent to the one introduced in Ref. \cite{intertwined}, Fig. S2. This is the globally most stable defect configuration that we identified.

In Fig. \ref{stmcomposite} we compare the measured and simulated STM images for this structure. Both filled and empty state images are in excellent agreement, indicating that the fused heart-shaped and In adatom PFDs indeed constitute the well-known short $\gamma$-PSD.

\section{Summary}

In summary, we determined the detailed atomistic structures of the most commonly observed phase flip and slip defects within the Si(111)-(8$\times$2)In atomic wire array using joint high resolution scanning tunneling microscopy measurements and \emph{first principles} calculations. We demonstrated that the well-known short phase flip defects can be caused \emph{either} by indium adatoms \emph{or} atomically-sized topological solitons. Structurally, the adatom-induced and solitonic phase flip defects are almost identical and closely resemble each other in scanning tunneling microscopy experiments. The adatom-induced and solitonic phase flip defects interact attractively. Their fusion is identically the known short phase slip defect. Our findings resolve the long-standing debate about the origin of the short phase flip and slip defects in this system. We propose the short atomically-sized left-chiral soliton as a highly useful model system to study topological electronic properties in quasi one-dimensional atomic wires.

\section{Acknowledgements}

  ASR and SW are supported by the German Federal Ministry of Education and Research (BMBF) within the NanoMatFutur programme (Grant No. 13N12972) and the German Research Foundation (Grant No. DFG FOR1700, WI3879/2). THK is supported by the  National Research Foundation of Korea (NRF) funded by the Ministry of Science and ICT, South Korea (Grant  No. 2018R1A5A6075964, 2016K1A4A4A01922028). HWY is supported by Institute for Basic Science (Grant No. IBS-R014-D1). Supercomputer time was provided by the Max-Planck Computing and Data Facility (MPCDF) Garching.

\end{document}